# LOW EMITTANCE LATTICE CELL WITH LARGE DYNAMIC APERTURE

A. Bogomyagkov, E. Levichev, P. Piminov, BINP, Novosibirsk 630090, Russia

*Abstract*

Compact low emittance lattice cell providing large dynamic aperture is essential for development of extremely low (pm range) emittance storage rings. As it is well known, a pair of identical sextupoles connected by a minus-identity matrix transformer in ideal case of kick-like magnets provides infinite dynamic aperture. Though the finite sextupole length degrades the aperture, it is still large enough, and in this report we discuss development of the low emittance lattice cell providing the –I condition for both horizontal and vertical chromatic sextupoles. Such cell can be used as a module for lattices of different emittance and length. As an example we develop a 3 GeV 10 pm emittance storage ring and study its transverse dynamic aperture.

## INTRODUCTION

In recent years several proposals of extremely low emittance storage rings were discussed (see, for instance, review in [1]). Lattices for an "ultimate light source" with emittance in the pm rad range providing fully diffraction-limited photon beams are being studied. It was revealed that one of the main constraints for such storage rings is reduction of dynamic aperture (DA) together with emittance. Qualitative explanation of this fact is simple: emittance minimization gives very low dispersion function outside of the bending magnet and strong chromatic sextupoles shrink DA. Quantitative estimation by the sextupole resonances fixed points study [2-4] scales emittance, dynamic aperture and integrated sextupole strength against the cell bending angle $\phi$ as

$$\varepsilon_x \propto \phi^3, \quad A_{x,y} \propto \sqrt{\varepsilon_x}/\xi_{x,y} \propto \sqrt{\phi^3},$$
$$(k_2 l)_{x,y} \propto \xi_{x,y}/\sqrt{\varepsilon_x} \propto 1/\sqrt{\phi^3},$$

where $\xi_{x,y}$ is natural chromaticity of the cell.

A widespread approach for DA increase exploits simultaneous optimization of dozen of parameters such as resonances driving terms of several orders, tune amplitude dependences, etc. by many families of sextupole and octupole magnets [5]. Besides this method does not guarantee success and requires long computing time, additional magnets occupy a lot of room at the ring, increase its length and, finally, the cost.

Meanwhile, arrangement of the chromatic sextupoles in pairs of identical magnets connected by –I transformer [6] eliminates all nonlinear aberrations, and one can achieve large DA (infinite for point-like sextupoles) independent of emittance and other lattice details. For the finite length sextupoles the second order aberrations are cancelled exactly while others remain, however even in this case dynamic aperture is still large enough for both beam injection and life time. Additionally, the third order aberrations (octupole-like) can be mitigated by a correction scheme described in [7], allowing dynamic aperture increase.

Below, we discuss a compact low emittance optical cell providing the –I condition in both planes. The cell can be used as a block to build storage ring with desired energy, emittance, circumference, straight sections number and length, etc. An option when the –I transform occurs in the one plane is also considered.

Finally, we demonstrate our approach in a 3 GeV storage ring structure with horizontal emittance of $\varepsilon_x = 10$ pm which is a diffraction limit for 1 Å radiation).

## CELL DESIGN PRINCIPLES

### Task definition

We start with the most appropriate for the low emittance cell TME lattice [8] depicted in Figure 1, left. As it will be shown later, in this cell we cannot adjust the –I transform for both transverse coordinates, so we modify it slightly by splitting the magnet and introducing the quadrupole q1 between two magnet halves (Figure 1, right). A Split Magnet TME (SM-TME) cell is still compact, provides low emittance but optically more flexible than the original TME.

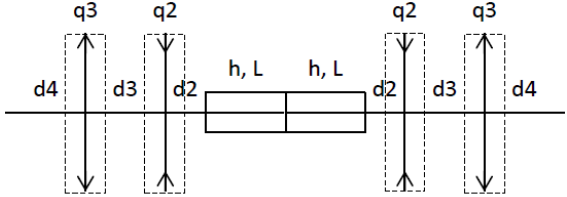 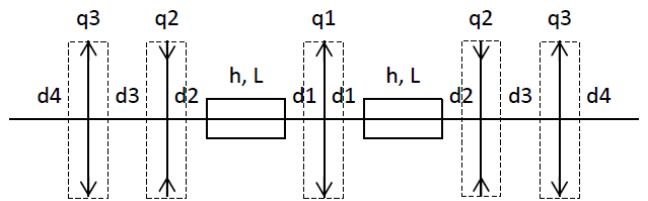

Figure 1: Scheme of the TME cell (left) and SM-TME (Split TME) cell (right).

In spite of q1 being denoted in Figure 1 as a focusing one, a solution with defocusing quad is also available and studied below.

Practical constraints we use below for numerical example include:

- The ring energy is 3 GeV.

- The magnet length (one-half of the original TME magnet) is L = 0.5–2 m; the quadrupole and sextupole length are ≤ 0.3 m. The drifts length small (for the sake of compactness) but reasonable from the viewpoint of location of the coils ends, flanges, BPMs and other standard accelerator components.
- Maximum quadrupole and sextupole strength are ~100 T/m and ~5000 T/m$^2$ respectively, which seems suitable for 25–30 mm poles bore diameter magnets.

To build the whole ring we follow a traditional approach proposed by D.Einfeld [9], when several TME cells compose the MBA (Multiple Bend Achromat) super-cell, bordered at both sides by dispersion suppressors. In our case we decided to have 45×5BA super-cells with the following bending angle formula: 45×(5×1.46°+0.7°) = 360°, where the regular and the dispersion suppressor magnet bends are 1.46° and 0.7° respectively.

The main goals of our study are as follows:
- Design of the SM-TME cell with the $-I_{x,y}$ condition for both horizontal and vertical sextupole pairs. As an option we also consider the $-I_x$ transform because usually the horizontal aperture is more important for machine performance than the vertical one.
- Minimization of the SM-TME cell emittance.
- Optimization of the lattice functions to get technically appropriate parameters of the cell magnets.

After development and detailed study of the Split TME structure we have recognized that similar cell was proposed earlier for KEKB arcs by H.Koiso and K.Oide [10] however they did investigate the cell chromatic performance and did not impose a low emittance condition combined with maximum cell compactness. So we believe that systematic exploration of the cell attractive for diffraction limited storage rings is, nevertheless, interesting and useful.

Below, $p_i$ denotes the normalized integrated strength (inverse focal length) of the quadrupole $q_i$ (see Figure 1): $p_i = (GL)_i / B\rho = -1/f_i$. The strategy of the cell design is as follows. First of all we construct the $-I_{x,y}$ transformer in the horizontal and vertical planes by the quadrupoles $q_2$ and $q_3$ for free ($p_1$, $L$) and fixed ($d_3$, $d_4$). To facilitate computations we use a simplified cell model shown in Figure 2. Here all the quads are considered as thin lenses and $L$ is equal to the magnet length with drifts separating it from the quadrupoles $q_1$ and $q_2$: $L = L_m + d_1 + d_2$. Since $d_1$ and $d_2$ are short with respect to the magnet length $L \approx L_m$. At the second step we minimize emittance by optimization of $p_1$ and $L$ in the $-I_{x,y}$ SM-TME cell. As an option, we consider relaxed condition of only horizontal $-I_x$ transformer and impose the vertical stability conditions to this particular cell. Finally, we apply our approach to the lattice cell design and study the dynamic aperture with the help of particle tracking.

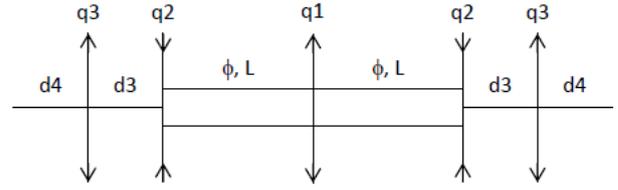

Figure 2: Simplified cell model.

### $-I_{x,y}$ transformer

The simplest conditions to cancel aberrations by two equal sextupoles are $\beta_{x,y1} = \beta_{x,y2}$, $\alpha_{x,y1} = \alpha_{x,y2}$, $\Delta\mu_{x1-2} = n_x \pi$, $\Delta\mu_{y1-2} = n_x \pi$, $n_{x,y} \in odd$. More sophisticated schemes can be found in [11].

Solutions $p_2 = f(p_1, L)$ and $p_3 = f(p_1, L)$ corresponding to $-I_{x,y}$ transformer for three example sets of ($d_3$, $d_4$) are demonstrated in Figure 3. Solutions have a reflection symmetry with respect to $p_1 = 0$. Crossing points of $p_2 = f(p_1, L)$ and $p_3 = f(p_1, L)$ show the case when $-I_{x,y}$ is achieved for the same $L$ and $p_1$.

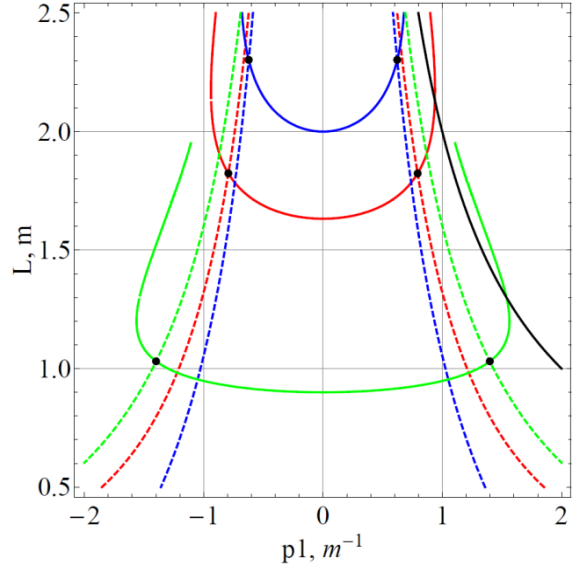

Figure 3: Dashed and solid lines refer to the solutions $p_2 = f(p_1, L)$ and $p_3 = f(p_1, L)$ corresponding to the $-I_{x,y}$ conditions over the cell. Colour indicates three sets of ($d_3$, $d_4$): green is for (0.5 m, 0.4 m), blue is for (1 m, 1 m) and red is for (1.13 m, 0.5 m). The black curve represents $L \cdot |p_1| = 2\phi \cdot \cot\phi$.

Analysis shows that the $-I_{x,y}$ solutions for different ($d_3$, $d_4$) (black points at Figure 3) roughly correspond to the following simple estimation
$$L \cdot |p_1| \approx 2\phi \cdot \cot\phi \approx (\phi << 1) \approx 2.$$

For the example we discuss below ($2\phi = 1.46°$) the exact product is $L \cdot |p_1| = 1.999$. Shorter magnet requires stronger $q_1$ but, fortunately, even for rather short magnets ($L < 0.5$ m) $|p_1|$ is still technically acceptable. Below we shall explore the following parameters $2\phi = 1.46°$,

$d_3 = 1.13$ m, $d_4 = 0.5$ m with the solutions providing $-I_{x,y}$ over the SM-TME cell equal to $L = 1.824$ m and $p_1 = \pm 0.79$ m$^{-1}$. For such inverse focal length the central quadrupole parameters are $l = 0.1$ m, $G = 80$ T/m at 3 GeV which satisfy well the modern small-aperture magnet technology.

Also one can see from Figure 3 that for the original TME cell ($p_1 = 0$) there are no solutions for the $-I_{x,y}$ transformer in both planes; however for one plane only (either horizontal or vertical) this solution exists.

*Emittance minimization*

A requirement for the $-I_x$ transformer in the horizontal plane unambiguously defines the dispersion function in the midpoint of the central quadrupole $q_1$:

$$\eta_1 = L \frac{(1-\cos\phi)}{\phi}. \qquad (1)$$

An intrinsic feature of the $-I$ cell is optical transparency: periodic lattice functions exist for any initial betas. So the horizontal beta profile can be chosen from the emittance minimization requirement only. For the horizontal beta $\beta_{x1}$ in the centre of $q_1$ the following emittance expression can be found

$$\varepsilon_x \approx C_q \gamma^2 \phi^3 \left[ \frac{2L}{15\beta_{x1}} + \beta_{x1}\left(\frac{1}{3L} - \frac{p_1}{8} + \frac{Lp_1^2}{80}\right) \right], \qquad (2)$$

where $\gamma$ is the relativistic factor and $C_q = 3.8319 \cdot 10^{-13}$ m. The emittance minimization with respect to $\beta_{x1}$ gives

$$\varepsilon_{x\min} = \frac{C_q \gamma^2 (2\phi)^3}{12\sqrt{15}} \cdot \frac{3}{2\sqrt{30}} \sqrt{5 + 3(Lp_1 - 5)^2} = \varepsilon_{TME} \cdot \frac{3}{2\sqrt{30}} \sqrt{5 + 3(Lp_1 - 5)^2} \qquad (3)$$

for

$$\beta_{x1\min} = \frac{4\sqrt{2}L}{80 - 30Lp_1 + 3L^2 p_1^2}. \qquad (4)$$

In above formula $\varepsilon_{TME}$ is the minimum emittance for the TME cell with $2\phi$ bending angle.

Figure 4 demonstrates the plot of equation (3) providing the following conclusions:

- For the product $L \cdot p_1 \approx 2$, corresponding to the $-I$ transformer in both planes, the SM-TME emittance is around 50% larger than the minimum one for the TME cell. Keeping in mind that the exact TME conditions require very strong optics that can hardly be met in real life, the results is not so bad.

- Another point $L \cdot p_1 \approx -2$, which also satisfy the $-I$ transformer in both planes, gives much larger emittance and is unacceptable.

- The emittance reduction below the TME minimum $\varepsilon_{SM}/\varepsilon_{TME} \approx 0.6$ is possible for the SM-TME cell but requires rather strong quadrupole q1 for reasonable magnet length $L \cdot p_1 = 5$. No $-I$ conditions exist in this case.

- The point $L \cdot p_1 = 0$ corresponding to the TME cell with $-I_x$ superimposed has emittance 2.5 times larger than the pure TME.

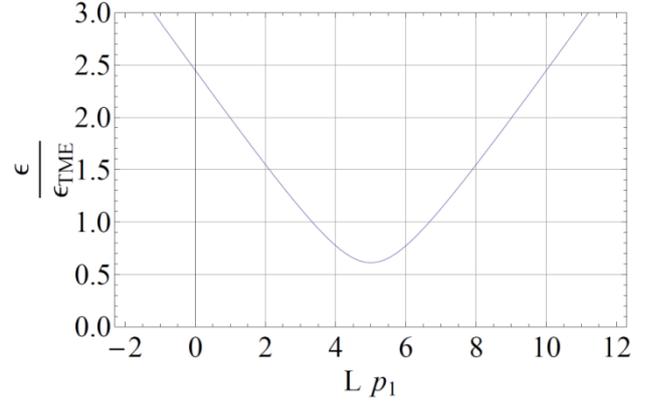

Figure 4: Ratio of the minimum emittances $\varepsilon_{SM}/\varepsilon_{TME}$ vs. $L \cdot p_1$.

*$-I_x$ transformer*

Sometimes large dynamic aperture is needed in one direction only (usually in horizontal). In this case the optical matching conditions are more relaxed, and we can consider, as before, $-I_x$ with the emittance minimization, but instead of $-I_y$ we only require the vertical stability, which here has usual meaning $-1 < \cos\mu_y < +1$.

Figure 5 shows such solutions at the $L(p_1)$ plot as shadowed areas for the positive and negative $p_1$. The dashed lines with boxed numbers are the contours of the ratio of SM-TME emittance and TME one (see eq. (3)). Two bold points show solutions corresponding to $-I_{x,y}$ transformer. The negative p1 provides, as it was mentioned already, larger emittance so we shall not consider these solutions below. In original TME cell with a single block magnet both conditions (vertical stability and $-I_x$ transformation) are available (see the line $p_1 = 0$ in Figure 5) but emittance is not so good ($\approx 2.5$ times of the TME minimum). The SM-TME emittance smaller than the TME unfortunately does not satisfy the vertical stability criteria and we failed to find optical functions corresponding to this region. For our purpose the most attractive area is shown in Figure 5 by the red box that is enlarged in Figure 6.

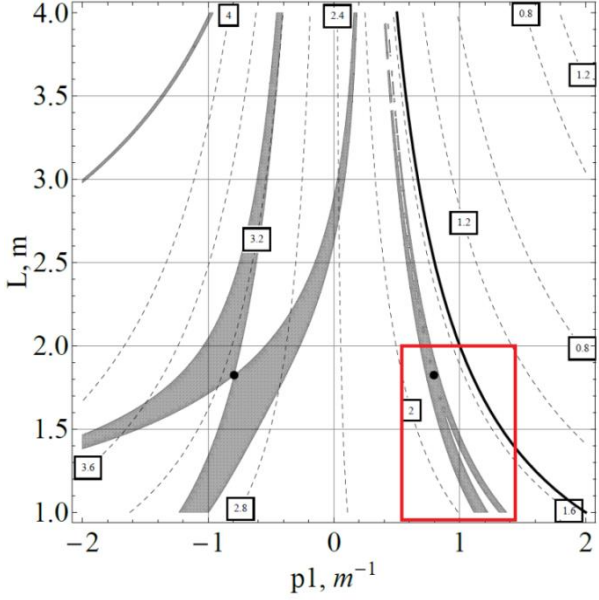

Figure 5: Vertically stable optical solutions (grey areas) with $-I_x$ condition over the SM-TME cell for the following parameters: $2\phi = 1.46°$, $d_3 = 1.13$ m, $d_4 = 0.5$ m.

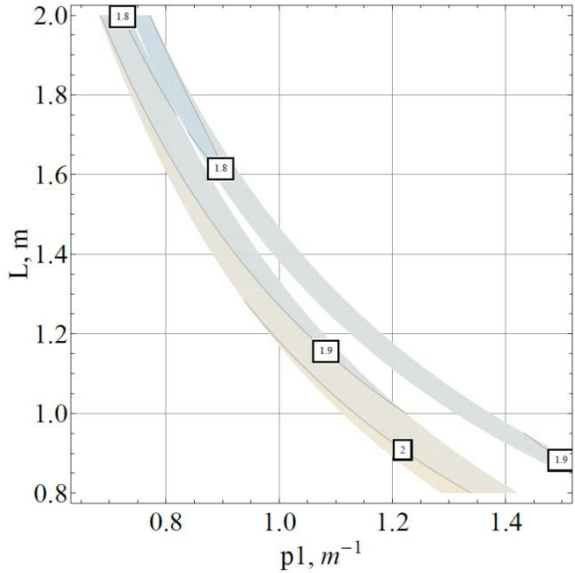

Figure 6: Solutions for the vertically stable SM-TME cell with $-I_x$ transformer.

The emittance of the cell is ≈1.8 larger than the TME minimum with the $2\phi = 1.46°$ dipole bend.

## TEST LATTICE

In this section we apply described approach to construct the lattice cell with 10 pm emittance and $-I_{x,y}$ transformation. Then we design a *super-cell* with 5BA SM-TME cells and dispersion suppressions at its both sides. We do not consider particular straight sections between the super-cells because their details depend on what kind of equipment (wigglers, undulators, RF cavities, injection magnets, etc.) they will accommodate. Instead we shall insert the general Twiss matrix between the super-cells and varying the matrix phase advances study which ones provide the maximum DA. These values can be used as recommendations for design of the real straight sections. We refer to the super-cell with two dispersion suppressors and mockup straight section represented by the matrix transformation as a ring period. Forty five periods compose a whole storage ring.

*Lattice cell*

Emittance minimization together with $-I$ transformation in both directions defines dispersion function in the cell in the unique way (1). Initial beta functions are defined by two factors: minimum emittance (horizontal beta) and minimum strength of chromatic sextupoles (both betas).

For our example parameters ($2\phi = 1.46°$, $d_3 = 1.13$ m, $d_4 = 0.5$ m, $L = 1.824$ m and $p_1 = 0.79$ m$^{-1}$) horizontal emittance as a function of initial $\beta_{x0}$ is shown in Figure 7. Minimum of the curve corresponds to the central beta value (4), however the curve around the minimum is rather flat, therefore initial $\beta_{x0} \approx 10 \div 25$ m is feasible providing cell matching flexibility.

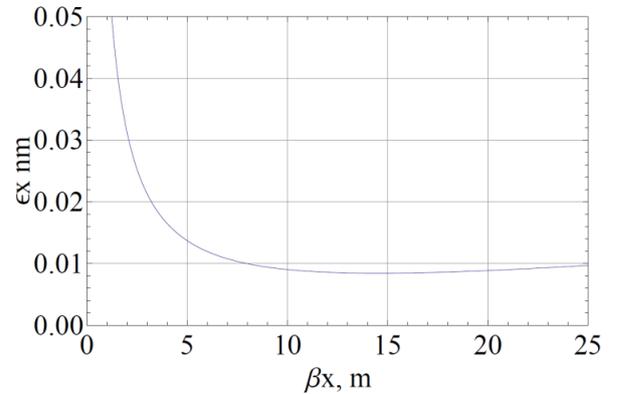

Figure 7: The cell emittance vs. the initial horizontal beta.

Figure 8 shows the cell chromaticities as functions of the cell initial betas with minimums at $\beta_{x0} \approx 4$ m and $\beta_{y0} \approx 1.2$ m. To compensate the natural chromaticity we installed sextupoles close to $q_3$ (horizontal) and $q_2$ (vertical) and found their integrated strengths as functions of the initial betas values. Results are given in Figure 9. Horizontal sextupole strength (S1) is small in the range of the beta values providing minimum emittance, while vertical sextupole strength (S2) is low for very small $\beta_{y0}$. Note that the minimum of the sextupole strength does not coincide with the minimum of chromaticity.

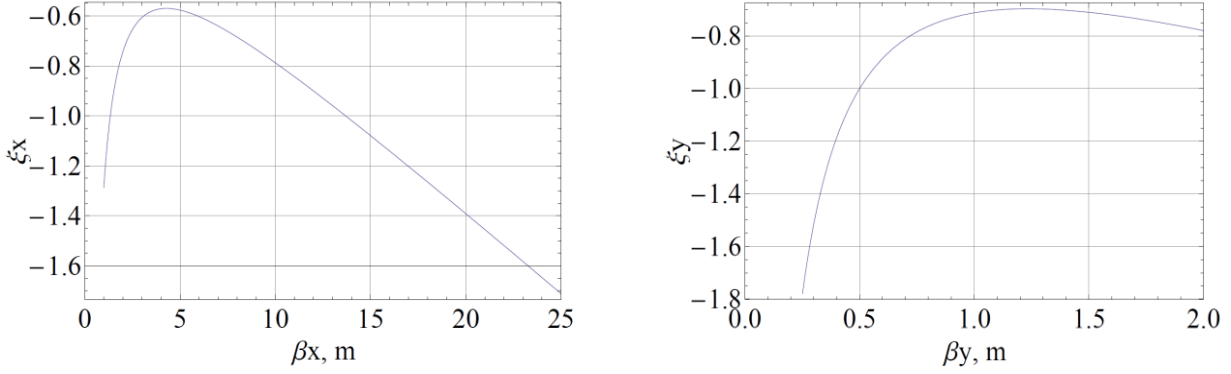

Figure 8: Cell chromaticity as a function of the corresponding initial beta.

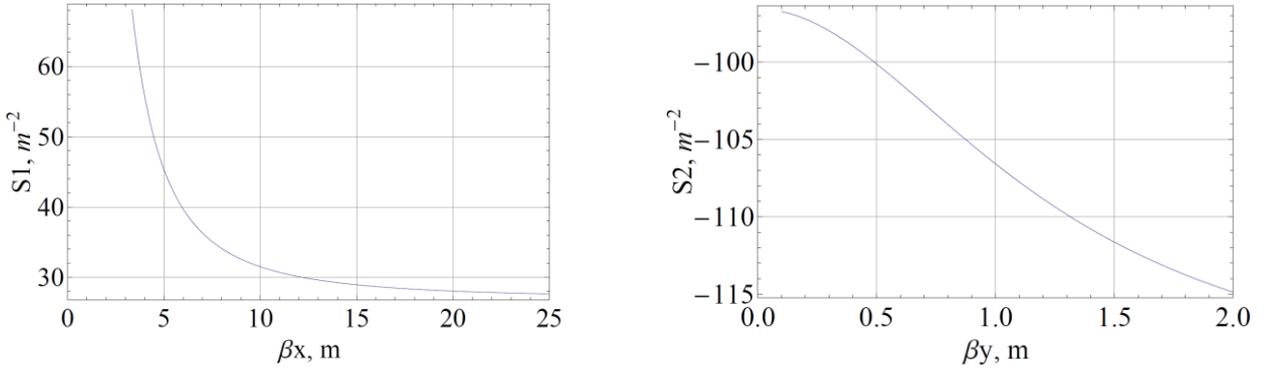

Figure 9: Sextupole strength needed to correct natural cell chromaticity as a function of initial betas.

Vertical sextupole strength is around 3 times larger than the horizontal one, however even the maximum integrated strength $(B''l)_y / B\rho \approx -100$ m$^{-2}$ gives for 3 GeV and $l = 0.2$ m the sextupole gradient $B'' = 5000$ T/m$^2$ that is not too large for the aperture 20÷30 mm. Finally Figure 10 shows the lattice cell magnets and optical functions.

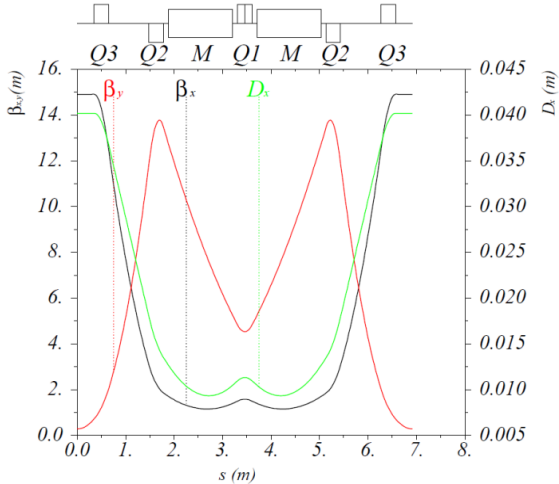

Figure 10: Split magnet TME cell for ultimate storage ring.

The cell possesses mirror point symmetry. The sequence of elements for the half of the cell is listed in Table 1. Q1 is the central quadrupole in the split magnet.

Table 1: List of the half-cell elements at 3 GeV

| Type  | Name | L[m] | B[T]/$\phi°$ | K1[m$^{-2}$]/B'[T/m] |
|-------|------|------|--------------|---------------------|
| Drift | O4   | 0.35 |              |                     |
| Quad  | Q3   | 0.3  |              | 1.93/19.3           |
| Drift | O3   | 0.83 |              |                     |
| Quad  | Q2   | 0.3  |              | -3.15/-31.5         |
| Drift | O2   | 0.1  |              |                     |
| Bend  | M    | 1.32 | 0.096/0.73   | 0                   |
| Drift | O1   | 0.1  |              |                     |
| Quad  | Q1   | 0.3  |              | 2.80/28.0           |

*Lattice super-cell structure*

Described cell provides conditions for both horizontal and vertical $-I$ transformers but can accommodate only one sextupole pair. To correct chromaticity in both planes we need several cells with horizontal and vertical sextupole pairs respectively. Below we use 5BA super-cell symmetrical around the midpoint with two chromatic correction sections in each direction as it is shown in Figure 11.

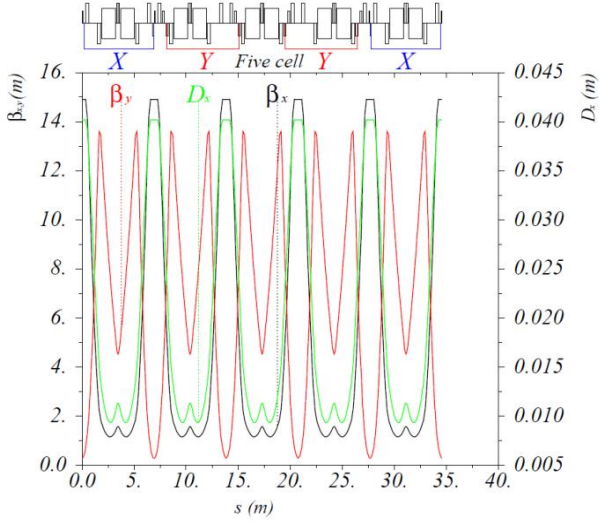

Figure 11: Five-cell superperiod with two horizontal sextupole pairs (denoted X) and two vertical ones (denoted Y).

The 5BA super-period has dispersion suppressors at both ends to close the ring geometrically. The suppressors are trivial and therefore not shown in Figure 11. To complete the ring design we need a number of straight sections which connect the super-cells and accommodate insertion devices, RF cavities or injection magnets. Instead of constructing particular straight sections which length and lattice functions depend on type of equipment they will accommodate, we connected the super-cells by Twiss matrix with phase advances, which can be adjusted to get the maximum DA. Main parameters of the cell, 5BA, super-cell and the whole ring (45 super-cells with couple of dispersion suppressors and dummy straight each) are listed in Table 2.

Table 2: Main parameters of the lattice at 3 GeV

| Parameter | SM-TME cell | 5BA | Super-cell | Ring |
|---|---|---|---|---|
| Length [m] | 6.92 | 34.59 | 54.04 | 2431.86 |
| Phase advance $\mu_{x,y}/2\pi$ | 0.5<br>0.5 | 2.5<br>2.5 | 3.463<br>3.418 | 155.85<br>153.82 |
| Emittance $\varepsilon_x$[pm] | 10.48 | 10.48 | 10.13 | 10.13 |
| Damping times $\tau_x/\tau_e$[ms] | 0.49<br>0.25 | 0.49<br>0.25 | 0.74<br>0.37 | 0.74<br>0.37 |
| Energy spread $10^4 \times \sigma_E/E$ | 2.5 | 2.5 | 2.5 | 2.5 |
| Natural chromaticity $\xi_x/\xi_y$ | -1.05<br>-1.57 | -5.25<br>-7.87 | -7.05<br>-9.76 | -317.24<br>-439.40 |

## Dynamic aperture

For kick-like sextupoles arranged in the $-I$ pairs all geometrical aberrations are cancelled exactly. For the real length magnets the second order aberrations (sextupole) are vanished but higher orders remain and cause dynamic aperture degradation. The leading perturbation order is the third one but map polynomial terms differ from those produced by octupole. Additional mitigation for the third order effects is possible (see, for instance, [7], where low strength sextupole correctors increase DA by ~30-50%), however here we intend to make a point of the bare $-I$ chromatic section advantages and do not apply any correction schemes. Two families of the chromatic sextupoles are listed in Table 3.

Table 3: Sextupole strength at 3 GeV.

| Sextupole type | Number per super-cell | Length [m] | $B''$[T/m$^2$] |
|---|---|---|---|
| Horizontal | 4 | 0.15 | 4450 |
| Vertical | 4 | 0.3 | -5350 |

With SM-TME cell (and 5BA structure) parameters fixed the only knob affecting DA is period fractional tune. We have varied the period phase advances by $2\pi$ with the help of the matrix, connecting two adjacent super-cells and tracked transverse DA in usual way for 1000 turns. As we used the matrix, the observation point betas were fixed as $\beta_{x,y} = 10$ m for the whole tune plane. The tracking results are given in Figure 12, where a color scale shows horizontal/vertical DA size at the period tune plane.

The plot in Figure 12 clearly demonstrates that all third order resonances are effectively suppressed by the sextupole pairs and only the fourth order resonances (octupole-like) are visible. The largest DA is in the corner below the half-integer resonances $\nu_{xp} = n/2$ and $\nu_{yp} = m/2$. For the further study we fixed $\nu_{xp} = 3.463$ and $\nu_{yp} = 3.418$ which gave the lattice tunes $\nu_x = 155.85$ and $\nu_y = 153.82$. All first order amplitude-detuning coefficients are negative

$$\Delta\nu_{xp} \approx -70 \cdot J_x[cm] - 8 \cdot J_y[cm],$$

$$\Delta\nu_{yp} \approx -40 \cdot J_x[cm] - 15 \cdot J_y[cm],$$

where $J$ is the action, so the particle tunes move out of the half-integer resonance with the oscillation amplitude increase and this fact has a positive impact on the dynamic aperture size.

Figure 13 shows the transverse dynamic aperture for the chosen tune point; the aperture exceeds $\pm 2$ cm horizontally and 3 cm vertically without any additional optimization.

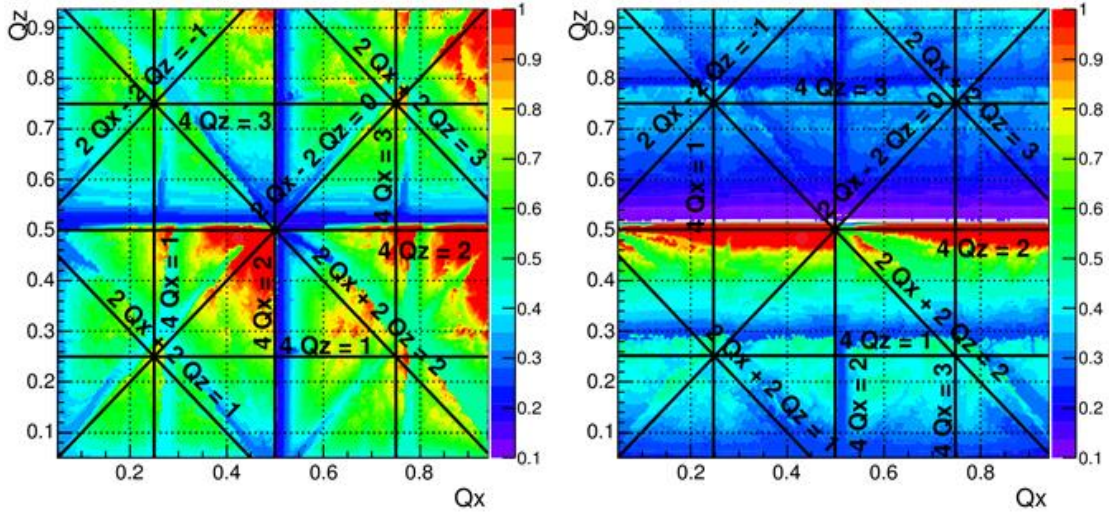

Figure 12: Horizontal (left) and vertical normalized DA a function of the period fractional tunes.

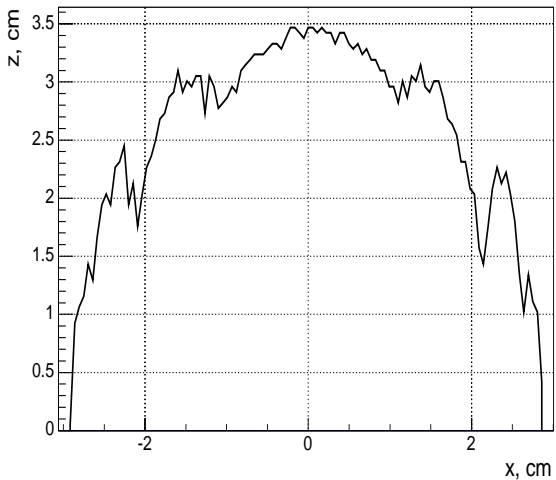

Figure 13: Dynamic aperture of the 10 pm emittance storage ring at $\beta_{x,y} = 10$ m.

## CONCLUSION

Compact Split TME lattice cell with low emittance providing the –I transformation in both planes is proposed and studied in details. Alternating identical cells with either horizontal or vertical sextupole pairs, one can compensate natural chromaticity in both planes.

The test storage ring based on the SM-TME cell is constructed with the horizontal emittance of 10 pm (which is a diffraction limit at 3 GeV beam energy). The –I sextupole pairs correcting chromaticity provide large dynamic aperture (more than ±20 mm horizontally and ±30 mm vertically at $\beta_{x,y} = 10$ m). The approach can be useful to design synchrotron light source, damping ring or other circular accelerator with small emittance. Additional sextupole or/and octupole correctors can increase the DA even more.